\documentclass[11pt]{article}
\usepackage{graphicx}

\newcommand{\BABARPubYear}    {04}
\newcommand{\BABARConfNumber} {007}
\newcommand{\SLACPubNumber} {10626}

\RequirePackage{xspace}

\usepackage{relsize}
\def\babar{\mbox{\slshape B\kern-0.1em{\smaller A}\kern-0.1em
    B\kern-0.1em{\smaller A\kern-0.2em R}}}
\def\Bbar    {\kern 0.18em\overline{\kern -0.18em B}{}\xspace}
\def\BB      {\ensuremath{B\Bbar}\xspace} 
\def\jpsi     {\ensuremath{{J\mskip -3mu/\mskip -2mu\psi\mskip 2mu}}\xspace}
\mathchardef\Upsilon="7107
\def\Y#1S{\ensuremath{\Upsilon{(#1S)}}\xspace}
\def\FourS {\Y4S}
\def\BR         {{\ensuremath{\cal B}\xspace}}
\newcommand{\gev}{\ensuremath{\mathrm{\,Ge\kern -0.1em V}}\xspace}
\newcommand{\mev}{\ensuremath{\mathrm{\,Me\kern -0.1em V}}\xspace}
\newcommand{\gevc}{\ensuremath{{\mathrm{\,Ge\kern -0.1em V\!/}c}}\xspace}
\newcommand{\mevc}{\ensuremath{{\mathrm{\,Me\kern -0.1em V\!/}c}}\xspace}
\newcommand{\gevcc}{\ensuremath{{\mathrm{\,Ge\kern -0.1em V\!/}c^2}}\xspace}
\def\ra                 {\ensuremath{\rightarrow}\xspace}
\def\to                 {\ensuremath{\rightarrow}\xspace}
\def\pep2{PEP-II}
\def\gsim{{~\raise.15em\hbox{$>$}\kern-.85em
          \lower.35em\hbox{$\sim$}~}\xspace}
\def\lsim{{~\raise.15em\hbox{$<$}\kern-.85em
          \lower.35em\hbox{$\sim$}~}\xspace}

\long\def\inst#1{\par\nobreak\kern 4pt\nobreak
    {\it #1}\par\vskip 10pt plus 3pt minus 3pt}

\begin{document}
{\pagestyle{empty}

\begin{flushright}
\babar-CONF-\BABARPubYear/\BABARConfNumber \\
SLAC-PUB-\SLACPubNumber \\
July 2004 \\
\end{flushright}

\par\vskip 3cm

\begin{center}
\Large \bf Measurement of the Inclusive Electron Spectrum in Charmless
Semileptonic {\boldmath $B$} Decays Near the Kinematic Endpoint and Determination of 
{\boldmath $|V_{ub}|$}
\end{center}
\bigskip

\begin{center}
\large The \babar\ Collaboration\\
\mbox{ }\\
\today
\end{center}
\bigskip \bigskip

\begin{center}
\large \bf Abstract
\end{center}
We study the inclusive electron spectrum of 
$B\to X_u e \nu$ decays near 
the kinematic limit for $B\to X_c e \nu$ transitions using a  
sample of 88 million \FourS\ decays 
recorded by the \babar\ detector 
at the PEP II $e^+ e^-$- storage rings.
For the electron momentum interval of $2.0 - 2.6 \gevc$ 
the partial branching fraction is measured to be 
$\Delta {\cal B}(B\ra X_u e \nu) =
(0.480 \pm 0.029_{stat} \pm 0.053_{syst})\times 10^{-3}$. 
Combining  this result on $\Delta {\cal B}$ with measurements of the inclusive 
photon spectrum in $B\ra X_s \gamma$ transition we find
$|V_{ub}|= 
(3.94 \pm 0.25_{exp} \pm 0.37_{f_u} \pm 0.19_{theory}) \times 10^{-3}$, 
where the first error represents  the combined statistical and systematic 
experimental uncertainty of the partial branching fraction measurement,
the second refers to the uncertainty of the determination of the fraction 
$f_u$ of the inclusive electron spectrum that falls within our cuts, and the third error is 
due to theoretical uncertainties in the QCD corrections, plus 
the uncertainty in the $b$-quark mass. All results are preliminary.\vfill
\begin{center}

Submitted to the 32$^{\rm nd}$ International Conference on High-Energy Physics, ICHEP 04,\\
16 August---22 August 2004, Beijing, China

\end{center}

\vspace{1.0cm}
\begin{center}
{\em Stanford Linear Accelerator Center, Stanford University, 
Stanford, CA 94309} \\ \vspace{0.1cm}\hrule\vspace{0.1cm}
Work supported in part by Department of Energy contract DE-AC03-76SF00515.
\end{center}

\newpage
}

\begin{center}
\small

The \babar\ Collaboration,
\bigskip

B.~Aubert,
R.~Barate,
D.~Boutigny,
F.~Couderc,
J.-M.~Gaillard,
A.~Hicheur,
Y.~Karyotakis,
J.~P.~Lees,
V.~Tisserand,
A.~Zghiche
\inst{Laboratoire de Physique des Particules, F-74941 Annecy-le-Vieux, France }
% ...

\end{center}\newpage

\section{Introduction}
\label{sec:introduction}

The increasingly precise measurements of CP asymmetries in $B$-meson
decays allow for stringent experimental tests of the Standard Model
mechanism for CP violation~\cite{sm} via the non-trivial phase in
the Cabibbo-Kobayashi-Maskawa  matrix. Improved measurements of
CKM element $|V_{ub}|$, the coupling of the $b$ quark to the $u$
quark, will enhance the sensitivity of such experimental tests. 

The extraction of $|V_{ub}|$ is a challenge, both theoretically and 
experimentally.  
Experimentally, the principal challenge is to separate the  signal $B \rightarrow
X_u e\nu$ decays from the 50 times larger $B\rightarrow X_c e\nu$ background.   
This is achieved by selecting regions of phase space in which the background is highly suppressed.  
In this paper we present a measurement of the inclusive electron spectrum
for charmless semileptonic $B$ decays, averaged over charged
and neutral $B$ mesons, near the kinematic endpoint.
In the rest frame of the $B$ meson, the
kinematic endpoint of the electron spectrum for the dominant 
$B \rightarrow X_c e \nu$ decays is $\sim2.3 \gevc$ and 
$\sim 2.6 \gevc$ for $B \rightarrow X_u e \nu$ decays. 
The finite momentum of the $B$ mesons in the $\Upsilon(4S)$ decays 
causes additional
spread of the electron momenta of $\sim 200$\mevc, extending the 
endpoints to higher momenta. A narrow interval of
about 300\mevc remains dominated by electrons
from $B \rightarrow X_u e \nu$ transitions, and this allows for
a relatively precise measurement of the partial branching fractions
in an interval that covers
approximately 10--15\% of the total electron spectrum for charmless
semileptonic $B$ decays free from significant \BB\ background. In
this analysis we extend the interval for signal extraction up to 
600\mevc\ covering $\sim 30\%$ of the total electron spectrum. 

Theoretically the weak decay rate for $b\ra u \ell \nu$ can be easily calculated
at the parton level. It is proportional to $|V_{ub}|^2$  and $m_b^5$, where
$m_b$ refers to the $b$ quark mass. To relate the $B$ meson decay rate to
$|V_{ub}|$, the parton-level calculations have to be corrected for perturbative
and non-perturbative QCD effects. These corrections can be calculated using
operator-product expansions in powers of $1/m_b$ and $\alpha_s$~\cite{ope}.
However, near the kinematic endpoint in the lepton spectrum, these calculations
break down because the spectra are affected by the ``shape function'', {\it i.e.,}
the distribution of the $b$-quark momentum inside the $B$ meson~\cite{neubert94},
in addition to weak annihilation and other effects. 
Thus extrapolation from the limited momentum range near the endpoint
to the full spectrum is a difficult task.
At present, the shape function cannot be calculated, but it should be
a universal property of the $B$ meson. To leading order it must
be the same for all
$b \ra q $ transitions (here $q$ represents any light quark).
It has been proposed~\cite{rothstein00}, \cite{neubert01}
that $|V_{ub}|$ can be extracted  by  comparing the high-energy end of the lepton
spectrum with the high end of the photon spectrum in $b\ra s \gamma$
transitions, thus reducing the theoretical uncertainty on the shape function.
Inclusive measurements of the photon spectrum are very challenging, but
two results are now available~\cite{cleo_photons}, \cite{belle_photons}.
This analysis is based on the same method as previous measurements 
of the lepton spectrum near the endpoint 
\cite{argus} (ARGUS), \cite{cleo}, \cite{cleo2}(CLEO).
The measurement of the partial branching fraction for charmless semileptonic
decays and the extraction of $|V_{ub}|$ presented here are updates of
preliminary results~\cite{ichep02} based on one-fourth the data.

\section{Data Sample, Detector, and Simulation}
\label{sec:Detector}

The data used in this analysis were recorded with the 
\babar\ detector at the \pep2 energy-asymmetric $e^+e^-$ collider.
The data sample of 88.4 million \BB\ events,
corresponding to an integrated luminosity 
of 80.4~$\mathrm{fb}^{-1}$,  was collected at the \FourS\ resonance;
an additional sample of 9.5~$\mathrm{fb}^{-1}$ was recorded 
at a center-of-mass energy just below the \BB\ threshold. 
The second data set is used for subtraction of the non-\BB\ contributions
from the data collected on the \FourS\ resonance.
The relative normalization of the two data samples
has been derived from luminosity 
measurements, which are based on the number of detected $\mu^+\mu^-$
pairs and the QED cross section for 
$e^+e^-\to \mu^+\mu^-$ 
production, adjusted for the small difference in center-of-mass energy. 

The \babar\ detector has been described in detail elsewhere \cite{detector}.
The most important components for this study are the charged-particle 
tracking system, consisting of a five-layer 
silicon detector and a 40-layer drift chamber, and the electromagnetic 
calorimeter assembled from 
6580 CsI(Tl) crystals. These detector components are embedded in a
$1.5\,\mathrm{T}$ solenoidal magnetic field. Electron candidates are selected on the 
basis of the ratio of the energy detected in the calorimeter 
to the track momentum, the calorimeter shower shape, 
the energy loss in the drift chamber, and the angle reconstructed 
in the ring-imaging Cherenkov detector.  

The electron identification efficiency and the probabilities to misidentify 
a pion, kaon, or proton as an electron have been measured as a function of
laboratory momentum and angles~\cite{thorsten} with clean samples of tracks 
that were selected from data. This experimental information is 
used in the Monte Carlo simulation to improve the agreement with the 
data. Within the acceptance of the calorimeter, defined by the polar angle
in the laboratory frame, $-0.72 < \cos \theta_{lab} < 0.92$,  the average
electron efficiency is $92\%$.  The average hadron misidentification rate
is about 0.1\%.

Tracking efficiencies and resolution have been studied in great detail. Comparisons 
with the simulation have revealed small differences, which have been taken 
into account. No significant impact of non-Gaussian resolution tails 
has been found for high momentum tracks contributing to the endpoint region.

We use Monte Carlo simulation of \BB\ events to estimate signal efficiencies
and background distributions. Most of the branching fractions for hadronic
$B$ and $D$ decays and form factors are based on values reported in the
Review of Particle Physics~\cite{pdg2002}.  
Charmless semileptonic decays, $B \ra X_u \ell \nu$, are simulated as a
mixture of three-body decays $(X_u = \pi, \eta, \rho, \omega,...)$
based on the ISGW2 model~\cite{isgw2}, and decays to hadronic states $X_u$,
with masses $m_{X_u}$ extending from $2 m_{\pi}$ to about 3.5 \gevcc, according
to the prescription of Ref.~\cite{dFN}.  The hadronization of $X_u$ is performed by
JETSET~\cite{jetset}. The motion of the $b$ quark inside the $B$ meson is
implemented with the shape function parameterization given in~\cite{dFN}.
The low-mass resonant decays are mixed with the non-resonant states in such
a way as to retain the cumulative distribution in the hadron mass $m_{X_u}$
as predicted by the non-resonant model of \cite{dFN}. 

For the simulation of the  dominant $B \ra X_c \ell \nu$ decays we have chosen
a variety of models. 
We use ISGW2~\cite{isgw2} for $B \ra D \ell \nu$ and the various decays to
higher mass $D^{**}$ resonances, adopt a prescription by Goity-Roberts~\cite{gr}
for the non-resonant $B \ra D^{(*)} \pi \ell \nu$ decays, and use an HQET
parameterization~\cite{hqet} of the form factors for $B \ra D^* \ell \nu$.
For these form factors we use the recent results from \babar\ ~\cite{babarff},
which have much smaller uncertainties than the earlier
measurements by CLEO~\cite{cleoff}.
The branching fractions for individual  $B \ra X_c \ell \nu$ decays are
adjusted to match the data (see below).

The Monte Carlo simulations include radiative effects such as bremsstrahlung
in the detector material and QED initial and final state radiation.
Adjustments for small variation of the beam energy over time have
also been included.

\section{Analysis}
\label{sec:analysis}

\subsection{Event Selection}
\label{sec:Selection}

We select semileptonic $B$-decay events by requiring that there be
an electron with momentum  $p_e > 1.1 \gevc$ in the \FourS\ rest frame.  
Throughout this paper, all quantities are measured in the \FourS\ rest
frame unless it is specified otherwise. To reject electrons from the
decay $\jpsi \ra e^+ e^-$ we combine the electron candidate with any
second electron of opposite charge and reject the candidate if the
invariant mass of the pair falls in the interval
$3.00 < m_{ee} < 3.15 \gevcc$. 

To suppress background from non-\BB\ events, primarily low multiplicity
QED (including $\tau^+ \tau^-$ pairs) and $e^+e^- \ra q \bar{q}$
 processes, we veto events with less than four charged tracks.  We
also require that the ratio of the second to the zeroth Fox-Wolfram
moments, $R_2=H_2/H_0 $~\cite{foxw}, not exceed $0.5$. $R_2$ is
calculated including all detected charged particles and photons.
For events with an electron in the momentum interval
of 2.0 to 2.6~\gevc, these two criteria reduce the non-\BB\ background
by a factor of about 6, while they retain more than 80\% of the signal
events.

In semileptonic $B$ decays, the neutrino carries  sizable energy.
In events in which the only undetected particle is
this neutrino, its energy and direction can be inferred from the
missing momentum in the event, $p_{miss}$. We estimate $p_{miss}$
from the difference between the four-momentum
of the two colliding-beam particles and sum of the four-momenta of all
detected particles,  charged and neutral.
To improve the reconstruction of the missing momentum, we impose a
number of requirements on the charged and neutral particles.
Charged tracks are required to have a minimum transverse
momentum of 0.2 \gevc and a maximum momentum of 10~\gevc in the
laboratory frame. They are restricted in polar angle to 
$-0.82 < \cos \theta_{lab} < 0.92$ and should originate close to the
beam-beam interaction point.
The detected energy of an individual photon is required to exceed 30 \mev.
The selection of semileptonic decays can be greatly enhanced by requiring
for the missing momentum that $|\vec{p}_{miss}|$ exceed  $ 0.5 \gevc$,
and that $\vec{p}_{miss}$   point into the detector fiducial volume,
$-0.9 < \cos\theta_{miss} < 0.9$, thereby effectively reducing the impact
of particle losses close to the beams.  Furthermore,
since in semileptonic $B$ decays, the neutrino and the electron are
emitted preferentially in opposite directions, we require that the angle
$\Delta \alpha$ between these two particles fulfill the condition
$\cos \Delta \alpha <0.4$.  These criteria on the missing momentum reduce
the continuum background from QED processes and $e^+e^- \ra q \bar{q}$
production by an additional factor of 3, while the signal loss is less than 20\%.

The detection efficiencies are estimated by Monte Carlo simulation. With the
stated selection criteria the efficiency (including effects of bremsstrahlung)
for detecting $B \ra X_u e \nu$ decays 
is close to 50\% and largely independent of the electron momentum between
1.1 and 2.0 \gevc; it gradually decreases above 2.0 \gevc,
reaching 35\% at 2.6 \gevc.

\subsection{Background Subtraction}
\label{sec:background}

\begin{figure}[!htb]
\begin{center}
\includegraphics[height=15cm]{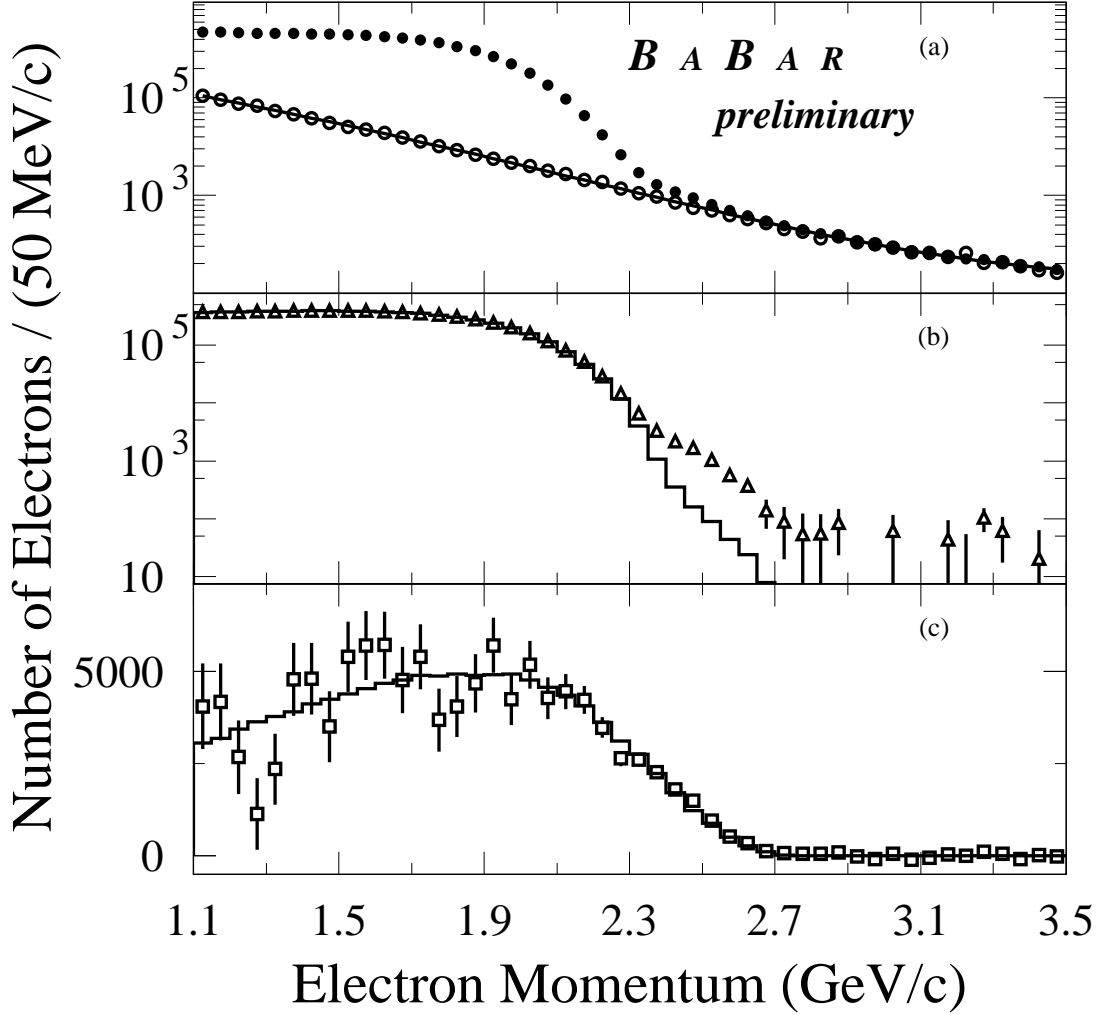}
\caption{
Electron momentum spectrum in the $\FourS$ 
rest frame: (a) on-resonance data (solid circles), 
scaled off-resonance data (open circles). 
The solid line shows the result of the fit 
to the non-\BB\ events
using both on- and off-resonance data in the 
interval $p_{e} = 1.1 - 3.5$\gevc;
(b) on-resonance data after subtraction of the fitted 
non-\BB\ background (open triangles) compared to simulated \BB\ background (histogram); 
(c) on-resonance data after subtraction of all backgrounds 
(open squares), compared to the simulated
$B \ra X_u e \nu$ events (histogram).
The error bars indicate statistical errors only.
}
\label{fig:p1}
\end{center}
\end{figure}

The spectrum of the highest momentum electron in events selected
by the criteria described above is shown in
Fig.~\ref{fig:p1}a, separately for data recorded on and
below the \FourS\ resonance.  
The data collected on the $\Upsilon(4S)$ resonance include
contributions from \BB\ events and non-\BB\ background. 
The latter is measured using off-resonance data, 
collected below \BB\ production threshold, and using on-resonance 
data above 2.8~\gevc, {\it i.e.,} above the endpoint for electrons 
from $B$ decays. The \BB\ background to the 
$B \to X_u e \nu$ spectrum is estimated 
from Monte Carlo simulation 
with the normalization of the individual contributions determined 
by a fit to the total observed spectrum. 

\begin{table}[h]
\caption{
Summary of sample composition: principal backgrounds and remaining signal in units of
$10^3$ events,
as well as the signal efficiencies. 
The errors are statistical, but for the non-\BB\ and $X_c e \nu$ backgrounds
the uncertainties in the fitted scale factors are included.}
\label{table:r1}
{\small
\begin{center}
\begin{tabular}{lr@{$\pm$}lr@{$\pm$}lr@{$\pm$}lr@{$\pm$}l }
 \hline \hline
\multicolumn{1}{l}{ $\Delta p\,(\mbox{GeV}/c)$ } & 
\multicolumn{1}{r@{$\div$}}{$2.0$} & \multicolumn{1}{l}{\hspace{-5pt}$2.6$}  &
\multicolumn{1}{r@{$\div$}}{$2.1$} & \multicolumn{1}{l}{\hspace{-5pt}$2.6$}  &
\multicolumn{1}{r@{$\div$}}{$2.2$} & \multicolumn{1}{l}{\hspace{-5pt}$2.6$}  &
\multicolumn{1}{r@{$\div$}}{$2.3$} & \multicolumn{1}{l}{\hspace{-5pt}$2.6$}  \\
\hline
                                                                                              
Total sample  & 609.81 & 0.78 & 295.76 & 0.54 &
              133.59 & 0.37 & 65.48 & 0.26 \\ \hline
Non-\BB\  & 142.33 & 0.63 & 105.15 & 0.48 &
              74.83 & 0.36 & 50.10 & 0.25 \\
$X_c e \nu$   & 423.25 & 1.58 & 160.17 & 0.86 &
              39.76 & 0.35 & 4.18 & 0.09 \\
$J/\psi$ and $\psi^{\prime}$ & 5.50 & 0.14 & 3.55 & 0.09 &
              2.07 & 0.06 & 1.04 & 0.03 \\
Other $e^{\pm}$     & 1.57 & 0.04 & 0.61 & 0.02 &
              0.23 & 0.01 & 0.07 & 0.01 \\
$\pi$ mis-ID  & 1.19 & 0.04 & 0.87 & 0.03 &
              0.57 & 0.02 & 0.31 & 0.02 \\
$K$ mis-ID    & 0.44 & 0.02 & 0.24 & 0.01 &
              0.12 & 0.01 & 0.05 & 0.01 \\
Other mis-ID  & 0.24 & 0.01 & 0.13 & 0.01 &
              0.07 & 0.01 & 0.03 & 0.01 \\
Other  $X_u e \nu$  & 1.33 & 0.07 & 0.54 & 0.03 &
              0.16 & 0.01 & 0.03 & 0.01 \\ \hline
$X_u e \nu$ signal & 33.97 & 1.92 & 24.50 & 1.17 &
              15.80 & 0.64 & 9.67 & 0.38 \\
\hline
$X_u e \nu$ efficiency (\%) & 43.2& 0.3 & 42.3 & 0.3 &
              41.3 & 0.4 & 40.4 & 0.5 \\
\hline \hline
\end{tabular}
\end{center}
}
\end{table}

\subsubsection{Non-\boldmath{\BB\ } Background}

To determine the non-\BB\ background we perform a $\chi^2$ fit to
the off-resonance  data, collected below $B\bar{B}$ production
threshold, and to  
on-resonance data in the momentum interval of 2.8~\gevc 
to 3.5~\gevc. Since the c.m. energy for the off-resonance data is
lower by about 0.4\%, we scale the lepton momenta by the energy
ratio to compensate for the difference.

The relative normalization for the two data sets is
$$ r_{L} = \frac{s_{\mathit{OFF}}} 
{s_{\mathit{ON}}} \frac{\int\! L_{\mathit{ON}}\,dt} 
{\int\! L_{\mathit{OFF}}\,dt} = 8.433\pm 0.004\pm 0.021, $$
where $s$ and $L$ refer to the c.m. energy squared and
luminosity of the two data sets.
The statistical uncertainty of $r_L$ is determined by the number of detected
$\mu^+\mu^-$ pairs used for the $\int\! L\,dt$ measurement; the 
systematic error of the ratio is estimated to be $0.25\%$.

The $\chi^2$ for the fit to the non-\BB\ events is defined as follows, 
\begin{equation}
\chi^2_c = \sum_i \frac{(f(\vec{a},p_i)-r_{L}  n_i)^2}{r_{L}^{2}  n_i}+
\sum_{j \hspace{0.1cm}(p_j>2.8\gevc)} \frac{(f(\vec{a},p_j)-N_j)^2}{ N_j}. 
\end{equation}

\noindent Here $N_j$ and $n_i$ refer to the number of selected events
on- and off-resonance 
in the $j$-th or $i$-th momentum bin, and
$\vec{a}$ is the set of free parameters of the fit. 
$f(\vec{a},p)$ is the function approximating the 
momentum spectrum, for which we have chosen an exponential expression of the form 
\begin{equation}
\label{f1} 
f(\vec{a},p) = a_1 + \exp(a_2 + a_3 p + a_4 p^2),
\end{equation} 
\noindent
The fit describes the data well, $\chi^2=70$ for 58 degrees of freedom.
Above 2.8 \gevc, we observe $(36.7 \pm 0.2)\times 10^3$ events in the on-resonance
 data, whereas the fit predicts $(36.6 \pm 0.2)\times 10^3$ events.

\subsubsection{\boldmath{\BB\ } Background}

The electron spectrum from $B$-meson decays  is composed 
of several contributions, dominated by  
the various semileptonic $B$-meson decays. Hadronic 
$B$ decays contribute mostly via hadron misidentification and
secondary electrons from decays of $D$, $J/\psi$, and $\psi^\prime$
mesons.

We estimate the total background by fitting the observed inclusive electron
spectrum to the sum of the signal and individual background contributions. 
For the individual signal and \BB\ background contributions, 
we use the Monte Carlo-simulated spectra and treat their 
relative normalization factors as free parameters in the fit.
The non-\BB\ background is parameterized by the exponential function
$f(\vec{a},p_i)$, as described above.
We expand the $\chi^2$ definition as follows, 
\begin{equation}
\chi^2 = \sum_i \frac{(f(\vec{a},p_i)-r_{L} n_i)^2}{r_{L}^2 n_i} +
           \sum_j \frac{(f(\vec{a},p_j)+S(\vec{b},p_j)-N_j)^2}{N_j + \sigma^2_{j\,MC}}, 
\end{equation}
\noindent where the first sum is for the off-resonance data 
and the second sum for the on-resonance data. The \BB\ electron 
spectrum is approximated as 
$S(\vec{b},p_j)=\sum_k b_k g_k(p_j)$, where the six free parameters 
$b_k$ are the correction factors to the default 
branching fractions for the individual contributions $g_k(p_j)$ representing the
signal $B\to X_ue\nu$ decays, the background $B\ra De\nu$, $B\ra D^*e\nu$, $B\ra D^{**}e\nu$,
$B\ra D^{(*)}\pi e\nu$  decays,
and the sum of other backgrounds with either electrons from secondary decays or
misidentified hadrons. 
$\sigma_{j\,MC}$ is a statistical error of the number of simulated events
in the $j$-th bin. The momentum spectra $g_k(p_j)$ are histograms from Monte Carlo
simulations. For the largest contribution, the $B \ra D^* e \nu$ decay, we
have adjusted the spectrum using the recent \babar\ measurements of the form
factor parameters~\cite{babarff}.

\subsubsection{Fit to Inclusive Spectra}

The fit is performed for the electron-momentum range from 1.1 \gevc to 3.5 \gevc, 
in bins of 50\mevc.  The lower part of the spectrum determines 
the relative normalization of the various background contributions, allowing 
for an extrapolation into the endpoint region above 2.0 \gevc.   
To reduce a potential systematic bias from the assumed shape of the signal
spectrum, we combine the on-resonance data for the interval from
2.2 to 2.8 \gevc into a single bin. The lower limit of this bin is chosen
so as to retain the sensitivity to the steeply falling \BB\ background
distributions, while containing a large fraction of the signal events
in a region where the background is
low.  As the limit is lowered to 2.0~\gevc , the resulting error on the
background subtraction increases.

The results of the fit and the subtraction of the fitted non-\BB\ and
\BB\ backgrounds are shown in Fig.~\ref{fig:p1}, and compared to
Monte Carlo simulations.  Above 2.3 \gevc, the  non-\BB\ background
is dominant, while at low momenta the semileptonic \BB\ background
dominates.  Contributions from hadron misidentification are small,
varying from 6\% to 4\% as the electron momentum increases.
The fit has a $\chi^2$ of $100$ for 75 degrees of freedom.

Table~\ref{table:r1} shows a summary of the data, principal backgrounds
and the resulting signal. The errors are statistical, but for
the non-\BB\ and  $X_c e \nu$ background
they include the uncertainties of the fitted parameters.
The data are shown for four different signal regions, ranging in
width from 600 \mevc to 300 \mevc. 
We choose 2.6~\gevc\ as the upper limit of the signal regions
because at higher momenta the signal contributions are very small
compared to the non-\BB\ background. The number of signal events
in a given signal interval is taken as the excess of events above
the fitted background.

\section{Systematic Errors}
\label{systematics}

The principal systematic errors originate from the fits to the backgrounds,
due to the uncertainties in the simulated momentum spectra of the various
contributions.
The uncertainty in the event simulation and thereby the
impact of the event selection on the momentum dependence of the
efficiencies for signal and background are the experimental limitations
of the current analysis. In addition, there are smaller corrections to the
momentum spectra due to variations in the beam energies, and  radiative
effects.  A summary of the systematic errors 
is given in Table~\ref{table:t2} for four intervals in the electron momentum. 

\begin{table}[h]
\caption{
Summary of the relative systematic errors (\%) on the partial branching
fraction measurements for  $B \ra X_u e \nu$ decays.
}
\label{table:t2}
{\small
\begin{center}
\begin{tabular}{lcccc} \cline{2-5} \hline \hline
$\Delta p~\, \mathrm{(GeV}/c)$ &
   $2.0 \div 2.6$  &
   $2.1 \div 2.6$  &
   $2.2 \div 2.6$  &
   $2.3 \div 2.6$  \\
\hline
Electron identification    & $0.8$ & $0.8$ & $0.8$ & $0.8$ \\

Track finding efficiency    & $0.7$ & $0.7$ & $0.7$ & $0.7$ \\
                                                                                           
$ B \ra X_u e\nu$ spectrum   & $7.9$ & $5.3$ & $3.2$ & $1.9$ \\
                                                                                              
Event selection efficiency      & $6.4$ & $6.7$ & $6.3$ & $5.7$ \\
                                                                                              
Non-\BB\ background       & $2.1$ & $2.1$ & $2.4$ & $2.6$ \\
                                                                                              
$B \to D^* l\nu$ form factor       & $1.7$ & $2.6$ & $1.7$ & $0.8$ \\
                                                                                              
$B \to D l\nu$ form factor         & $0.3$ & $0.9$ & $0.8$ & $0.7$ \\
                                                                                              
$B \to D^{**} e \nu$ spectrum       & $2.3$ & $2.0$ & $2.6$ & $1.3$ \\
                                                                                              
$J/\psi$ and $\psi^{\prime}$ background      & $0.8$ & $0.7$ & $0.7$ & $0.5$ \\
                                                                                              
Other $e^{\pm}$ background               & $0.5$ & $0.2$ & $0.1$ & $0.1$ \\
                                                                                              
$\pi$ mis-ID background                 & $0.7$ & $0.7$  & $0.8$ & $0.7$ \\
                                                                                              
$K$ mis-D background                   & $0.4$ & $0.3$ & $0.2$ & $0.2$ \\
                                                                                              
Other hadron mis-ID background      & $0.2$ & $0.2$ & $0.1$ & $0.1$ \\
                                                                                              
$B \to X_u e \nu$ background             & $1.2$ & $0.7$ & $0.3$ & $0.1$ \\
                                                                                              
$B$ momentum                        & $1.1$ & $1.5$ & $1.3$ & $0.5$ \\
                                                                                              
$N_{B\bar{B}}$ normalization       & $1.1$ & $1.1$ & $1.1$ & $1.1$ \\
                                                                                              
Bremsstrahlung and FSR correction     & $0.8$ & $1.1$ & $1.2$ & $0.7$ \\
\hline
                                                                                              
Total Systematic Error             & $11.1$ & $9.8$ & $8.5$ & $7.0$ \\
                                                                                              
\hline \hline
\end{tabular}
\end{center}
}
\end{table}

\subsection{Detection and Simulation of \boldmath{$B \to X_u e \nu$} Decays}

The detection efficiency for 
$B \rightarrow X_u\ell\nu$ decays is determined by Monte Carlo simulation.
We include in the uncertainty of the signal spectrum not only the
uncertainty in simulation of the detector response,
but also the uncertainty in the simulation of the momentum and angular
distributions of the charged lepton, as well as the hadrons and neutrinos.

\subsubsection{Detector related uncertainties}
For a specific model of the signal decays there are three major factors
that determine the efficiency: the track reconstruction for the electron,
the electron identification, and losses due to the detector acceptance
and the event selection.

The uncertainty in the tracking efficiency has been studied in detail
and is estimated to be $\sim 0.7\%$ per track.
The average identification efficiency
for electrons above 1.0~\gevc is estimated to be 
91.5\% ~\cite{thorsten}, based on large samples of radiative Bhabha
events and two-photon interactions.
In \BB\ events the actual efficiencies are slightly lower due to 
higher track multiplicity. 
This difference decreases gradually from about 2.5\% at 1.0~\gevc to less than 0.8\% 
above 2.0~\gevc. 
An independent estimate of 0.6\% for this uncertainty was derived from
a comparison of the efficiency in data and simulation, 
for electrons from $J/\psi K^{(*)}$ decays.
We adopt an error of 2\% for the systematic uncertainty in the 
electron-identification efficiency at 1.0 \gevc, decreasing to 0.8\%  above 2.0~\gevc.
In addition, we take into account the impact of the momentum-dependent
uncertainty of electron identification efficiency on the observed electron 
spectrum for both signal and background (see below).

\subsubsection{Uncertainties in the signal spectrum}

The momentum distributions of the signal electrons are not precisely
known because many of the $B \rightarrow X_u\ell\nu$ decay modes are still
unobserved or poorly measured due to small event samples, and even
for observed ones the decay form factors are not
measured. For decays with low  mass charmless hadrons the simulation is based on ISGW2 model. 
For decays to higher mass, mostly non-resonant states we rely on the model of 
de Fazio and Neubert~\cite{dFN} and a fragmentation model~\cite{jetset}. 

To evaluate the sensitivity of the signal efficiency to the decay multiplicity and 
the angular and momentum distributions, 
we randomly vary the individual branching ratios for decays to resonant
and non-resonant charmless hadrons by 50\%, except for 
$B\to\pi\ell\nu$ and $B\to\rho\ell\nu$ which are currently measured to 
30\% and 25\%, respectively.  We observe changes of less than 3.0\% for
the whole spectrum, and less than 1.0\% in the signal yield above 2.3\gevc.

The systematic uncertainties inherent in the modeling of the signal decays to 
non-resonant hadronic states have been studied by varying the shape-function
parameters as determined from the measurement of the inclusive photon
spectrum~\cite{cleo_photons}.
These variations translate to changes of the signal lepton spectrum 
and thus impact the fit to the overall spectrum. The resulting changes in
the signal branching fraction have been evaluated, taking into account
the errors and correlation of the shape function parameters.

Not included in this estimate is the sensitivity 
to the event selection criteria, specifically those based on the variables
$R_2$ and $p_{miss}$. These criteria not only influence the signal, but more
so the background distributions, and they are discussed below.

\subsection{Non-\boldmath{\BB\ }  Background}

Systematic errors in the subtraction of the non-\BB background
could be introduced by the choice of
the fitting function describing the 
lepton spectrum and by the uncertainty in the relative normalization of the
on- and off-resonance data.

To assess the uncertainty in the shape of this background we have compared fits with 
different parameterizations of the fitting function. In addition to the exponential 
function described above, we have tried linear combinations of Chebyshev polynomials 
up to the fifth order. The resulting fits are equally consistent with the data.
The differences in the non-\BB\ background estimates between different
parameterizations are at a level of less than 0.5\%.

If relative normalization is treated as a free fit parameter, its deviation
from the value based on luminosity measurements is less than one standard
deviation, which is equal to 1.5\%. Thus for the normalization 
the more accurate value based on luminosity measurements is used.
As a systematic error of non-\BB\ background we take 0.5\% of this
background contribution which includes the
errors of normalization factor and background shape approximation.

\subsection{\boldmath{$B \to J/\psi X$} Background}

$J/\psi$ and $\psi'$ decays to lepton pairs are vetoed by a restriction 
on the di-lepton effective mass. However, this veto is only about 50\% efficient,
mostly because of acceptance losses.
The remaining, mostly single track background is estimated from simulation. 
We observe a difference of $5.0 \pm 2.7\%$ between 
the veto efficiency for lepton pairs in data and simulation, and thus assign
a 5\% error to the residual background. 
Since this background amounts to  $18\%$ (10\%) of the signal for 
$p_{e} > 2.0~\gevc $
($p_{e} > 2.3~\gevc $) for both differential and integrated spectra, 
the resulting uncertainty on the signal branching fraction 
is estimated to vary from  0.8\% to  0.5\%.

\subsection{\boldmath{\BB\ } Background}

A major concern in this analysis is the uncertainty in the estimate 
of the signal events from $B\to X_u \ell \nu$ decays
obtained from the fit of the sum of various background contributions
to the observed spectrum. 
The shapes of \BB\ backgrounds are derived from Monte Carlo simulation.
The branching fractions 
for exclusive semileptonic $B\to X_c\ell\nu$ decays
are currently not precisely known. Thus the lepton spectra from
$B\to X_c\ell\nu$ decays may differ from those of the simulation. For
this reason we have introduced scale factors in the fits to the
spectrum to adjust the relative normalization of the various contributions.
To test the sensitivity to the shape of the dominant contributions, we have varied 
the form factor for decays to $D^* e \nu$ and $D e \nu$, and changed the relative 
proportion of contributions from narrow and wide resonances to $D^{**} e \nu$ decays.

The differential decay rate for $B\to D^*\ell\nu$ 
can be described by three amplitudes, which depend on the three 
parameters: $\rho^2$, $R_1$, and $R_2$. Their measured values
are   $\rho^2=0.769 \pm 0.039 \pm 0.019 \pm 0.032$,
$R_1=1.328 \pm 0.055 \pm 0.025 \pm 0.025$, 
and $R_2=0.920 \pm 0.044 \pm 0.020 \pm 0.013$ \cite{babarff}.
The quoted errors are statistical uncertainties from the data
and Monte Carlo samples, and systematic errors, respectively.
The $B\to D\ell\nu$ differential
decay rate can be described 
by a single parameter $\rho_D^2$. To study the impact of form factor variations 
we reweight the Monte Carlo-simulated spectrum for a given decay mode with 
the relative change of the generator-level spectrum due to changes in
the form-factor parameters, and repeat the standard fit to the data.
From the observed changes in the signal rate as a function of the choice of 
the form-factor parameters for $D^* e \nu$ decays, we assess the systematic error 
on the signal rate by taking into account the measured form-factor parameters, 
their errors, and their covariance matrix~\cite{babarff}.  
For $D e \nu$ decays, we rely on a measurement by the CLEO Collaboration~\cite{cleoffd},
$F_D(w)/F_D(1) = 1 - \rho_D^2 (w-1)$, where $\rho_D^2 = 0.76 \pm 0.16 \pm 0.08$ 
is the linear slope of the $w$-dependence. The variable $w$ is the product
of the four-vector velocities of the $B$ and $D$ mesons and corresponds to
the relativistic boost of the $D$ meson in the $B$ rest frame.  To estimate
the impact of the uncertainty in $\rho_D^2$ we likewise compare the default fit results 
with fits performed with a reweighted lepton spectrum. In this case, we replace 
the default simulation based on the ISGW2 model with the simpler form factor 
formulation and vary the measured value of $\rho_D^2$ by one standard deviation. 
We adopt the mean shift of the signal rate as a systematic error.
 
To assess the impact of the poorly known branching fraction for various 
$D^{**} e \nu $ decay modes on the shape of the lepton spectrum
we have repeated the fit with 
the different relative branching fractions for the individual decays
 modes.  As long as we do not eliminate the decays to the two narrow
resonances, $D_1(2437)$  and $D_2(2459)$, we obtain reasonable results.
Specifically, if we eliminate the decays involving
the two wider resonances, $D_0(2308)$ and $D_1'(2460)$, the results
change by less than 3\%.  We adopt this change as the estimate of
the systematic error due to the uncertainty of decays to $D^{**}$ states.

Similarly, we vary the branching ratios for secondary electrons 
from semileptonic $D$ decays by 10\% and adopt the observed change
as a systematic error. 
There is a small background from events which contain 
a $B\to X_u e \nu$ decay but contribute to the background rather
than the signal. We estimate the uncertainty in this contribution to be 30\%.

For background from hadronic $B$ decays, the uncertainty in the spectrum 
is primarily due to the uncertainty in the momentum-dependent hadron 
misidentification. The uncertainties of misidentification
probabilities are estimated to be 20\%, 30\%, and 50\% 
for pions, kaons, and protons respectively. The uncertainty 
in the fractions of pions, kaons, and proton is taken as a
difference between simulated and observed charged particle spectra, which is 
about 5\% for pions and kaons, and 50\% for much smaller contribution from protons
and antiprotons. 
With these uncertainties in the hadron misidentification backgrounds, 
the fractional error in the  number of subtracted background events 
is $\sim 20$\% for pions, $\sim 30$\% for kaons, and $\sim 70$\% for protons.

\subsection{\label{AB_syst_study} 
{Uncertainty in the \boldmath{$B$} Meson Momentum Spectrum}}

The non-zero momentum of the $B$ mesons in the \FourS\ rest
frame affects the shape of the electron spectra near the endpoint.
To estimate the systematic error of the inclusive lepton spectra 
associated with the uncertainty of initial $B$ meson momentum spectrum 
we compare the simulated and measured energy spectra for fully
reconstructed charged $B$ mesons.  The widths of the energy distributions
agree well, but in some of the data sets we observed a shift in the
central value of up to 2.2 \mev relative to the simulation.  We correct
the simulation for these shifts, and assess the effect of the uncertainty
of 0.13 \mev\ in this shift.

\subsection{Bremsstrahlung and Radiative Corrections}
\label{Radiative}

For comparison with other experiments and with theoretical calculations the signal 
spectrum resulting from the fit is corrected for bremsstrahlung in the detector and 
for final-state radiation. Corrections for QED radiation in the decay process are 
simulated using PHOTOS~\cite{photos}. This simulation includes multiple-photon emission 
from the electron, but does not include electroweak corrections for quarks. The
accuracy of this simulation has been compared to analytical 
calculations performed to ${\cal O}(\alpha)$~\cite{photos}. Based on this comparison
we assign an uncertainty of 10 - 15\% to the PHOTOS correction, leading to
an uncertainty in the signal yield of about 1\%.

The uncertainty in the energy loss due to bremsstrahlung  is determined by the
uncertainty in the thickness of the detector material, estimated to be
$(0.0450 \pm 0.0014) X_0$ at normal incidence.  This thickness was verified
using Bhabha scattering as a function of the polar angle relative to the beam. 
The impact of this uncertainty on the signal rate was estimated by calculating
the impact of an additional $ 0.0014 X_0$ of material; it is of the order of 1\%.

\subsection{\label{SES}
{Sensitivity to the Event Selection}}  

We have checked the sensitivity of the fits to the lepton spectrum to changes 
in the event selection, as well as the momentum dependence of 
the electron selection efficiencies. 
These variations of the cuts change the signal efficiency and lead to large
variations up to $50\%$ in the size of the non-\BB\ background 
and up to $20\%$ in the $\BB$ background. 
Though some of the observed changes in signal yield  may already be covered by the form
factor and other variations, we conclude that these tests do reveal
significant  changes that have to be accounted for.
We interpret the observed changes as inadequacies in the simulation of the
Monte Carlo-simulated  lepton spectra and adopt the observed changes between
the default fits and the looser selection criteria as  systematic errors. 

The largest variation (5\%) is observed for changes in the restriction of
the Fox-Wolfram ratio $R_2$ from the default value of 0.5 to 0.4 and 0.6.
Other sizable variations are observed for changes in the absolute 
value and direction of the missing momentum vector.
$R_2$ and the missing momentum are quantities that are derived from the measured
momenta of all charged 
and neutral particles in the event and are therefore sensitive to even small
differences in data and the simulation.
We consider the chosen variations of the cut variables representative for the
error estimation, add the
observed changes in quadrature, and include them in
the overall systematic error.

\section{Results}
\label{results}

\subsection{Determination of Partial \boldmath{$B\to X_u e \nu$} Branching Fraction}

For a given interval $\Delta p$ in the electron momentum, we  calculate the inclusive
partial branching fraction $B\to X_u e \nu$  according to 
\begin{equation}
\Delta {\cal B} = \frac{N_{tot}(\Delta p)-N_{bg} (\Delta p) } 
                {2\epsilon(\Delta p) N_{\BB}} (1+\delta_{\mathit{rad}}(\Delta p)).
\end{equation}
 
\noindent 
Here $N_{tot}$  refers to the total number of 
electron candidates detected  in the on-resonance data and $N_{bg}$ refers to 
the total background, from non-\BB\ and \BB\ events, as determined from the fit to the spectrum. 
$\epsilon(\Delta p)$ is the total efficiency for detecting a signal electron from 
$B \rightarrow X_u e\nu$ decays (including bremsstrahlung in detector 
material), and $\delta_{\mathit{rad}}$ accounts for the 
distortion of the electron spectrum due to final-state radiation. 
This is a momentum-dependent correction, derived from the 
Monte Carlo simulation based on
PHOTOS~\cite{photos}. The total number of produced $B\bar{B}$ events 
is $N_{\BB} = (88.36\pm 0.02 \pm 0.97) \times 10^6$.

\begin{figure}[!htb]
\begin{center}
\includegraphics[height=10cm]{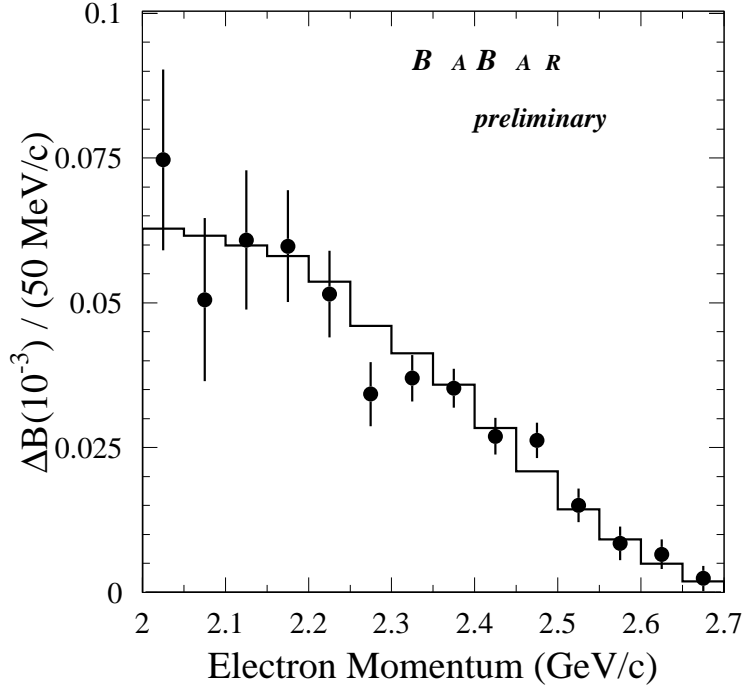}
\caption{
The differential branching fraction for charmless semileptonic $B$ decays (data points)
as a function of the electron momentum (in the $\FourS$ 
rest frame) after background subtraction and corrections for bremsstrahlung and
final state radiation, compared to the Monte Carlo simulation (histogram).
The errors indicate the statistical errors on the background subtraction,
including the uncertainties of the fit parameters.
}
\label{fig:p2}
\end{center}
\end{figure}
The differential branching fraction as a function of the electron momentum 
in the \FourS\ rest frame is shown in 
Fig.~\ref{fig:p2}, fully corrected for efficiencies and radiative effects. 
Partial branching fractions  for the four different momentum intervals are
summarized in Table~\ref{table:br2f}.  There is excellent agreement with
the preliminary \babar\  measurement~\cite{ichep02} and also with a
measurement by the CLEO Collaboration~\cite{cleo2}. 

\begin{table}[h]
\caption{
Preliminary results on the partial ($\Delta {\cal B}$) and the total branching
fraction (${\cal B}$) for inclusive $B \to X_u e \nu$ decays for four momentum intervals.
$f_u(\Delta p)$ is a fraction of the lepton spectrum, that falls into the
$\Delta p$ momentum interval. The values of $f_u(\Delta p)$ were derived by the CLEO 
Collaboration \cite{cleo2} from the shape
function parameters based on the measurement of the $b\to s\gamma$ spectrum.
}
\label{table:br2f}
{\small
\begin{center}
\begin{tabular}{ccccc} \hline\hline
$\Delta p$ (\gevc) & $\Delta {\cal B}\,(10^{-3})$ & $f_u(\Delta p)$ & ${\cal B}\,(10^{-3})$ & $|V_{ub}|\,( 10^{-3} )$ \\ \hline
$2.0 \div 2.6$  & $0.480\pm 0.029\pm 0.053$ & $0.278\pm 0.043\pm 0.030$ & $1.73\pm 0.22\pm 0.33$ & $3.94\pm 0.25\pm 0.37\pm 0.19$ \\ 
$2.1 \div 2.6$  & $0.355\pm 0.018\pm 0.035$ & $0.207\pm 0.037\pm 0.027$ & $1.71\pm 0.19\pm 0.38$ & $3.93\pm 0.22\pm 0.43\pm 0.19$ \\ 
$2.2 \div 2.6$  & $0.235\pm 0.010\pm 0.020$ & $0.137\pm 0.025\pm 0.023$ & $1.72\pm 0.16\pm 0.42$ & $3.93\pm 0.19\pm 0.48\pm 0.19$ \\ 
$2.3 \div 2.6$  & $0.149\pm 0.006\pm 0.010$ & $0.078\pm 0.015\pm 0.016$ & $1.90\pm 0.15\pm 0.52$ & $4.14\pm 0.16\pm 0.56\pm 0.20$ \\ 
\hline \hline
\end{tabular}
\end{center}
}
\end{table}

\subsection{Extraction of the Total Charmless Branching Fraction and \boldmath{$|V_{ub}|$}}

To determine the charmless semileptonic branching fraction 
${\cal B} (B\rightarrow X_u e\nu)$ from the partial branching fraction 
$\Delta {\cal B}(\Delta p)$, one needs to 
know the fraction $f_u(\Delta p)$ of the spectrum that falls into 
the momentum interval $\Delta p$.
The CLEO Collaboration \cite{cleo2} has used the measurement 
of the inclusive photon spectrum from $b\to s\gamma$ transitions to derive 
the parameters describing the shape function to determine the fraction 
$f_u(\Delta p)$, and to estimate the uncertainties of the resulting shape
function parameters. 
The $f_u(\Delta p)$ fractions are listed in Table~\ref{table:br2f},
together with the
total branching fractions. Contrary to the CLEO analysis, we have fully
corrected the $\Delta {\cal B}$ for radiative effects, and consequently
the fractions $f_u$ listed here do not include such corrections~\cite{gibbons}.
The quoted errors on ${\cal B}$ and $f_u$ are statistical and systematic.
The systematic error of $f_u$ includes the systematic error
of \BB\ subtraction, uncertainties of the shape function approximation,
dependence on $\alpha_s$ scale, and the uncertainty of theoretical prediction of $f_u$
from $b\to s\gamma$ shape function. The errors on $\cal B$ and 
$|V_{ub}|$ are the total error from $\Delta {\cal B}$ measurement
and the error from $f_u$. The last error in  $|V_{ub}|$ is
the theoretical uncertainty of the translation from $\cal B$ to $|V_{ub}|$.
Surprisingly, the overall precision does not depend very strongly on the
chosen momentum interval. While the experimental errors are smallest for
the interval from 2.3 to 2.6~\gevc, the dominant uncertainty arises from 
the determination of the fraction $f_u$ and this increases substantially
with higher momentum cut-off.  Thus we quote as the branching fraction
measurement the result based on the data in the interval $2.0 - 2.6$~\gevc,
\begin{equation}
\textrm{\BR}(B\rightarrow X_u e\nu)=
(1.73  \pm 0.22_{exp} \pm 0.33_{f_u})\times 10^{-3},
\end{equation}
where the first error represents the sum of the statistical and systematic
error on the partial branching fraction, and the second error refers to
the uncertainty in the measured photon spectrum and the extraction of the shape function. 
From the inclusive charmless semileptonic branching fraction 
and the average $B$ lifetime we can extract $V_{ub}$,
\begin{equation}
|V_{ub}| = 0.00424 \left( \frac{\textrm{\BR}(B \to X_u l \nu)}{0.002} 
\frac{1.604\,\mbox{ps}} {\tau_b} \right)^{1/2} 
(1.0 \pm 0.028_{OPE} \pm 0.039_{m_b}). 
\end{equation} 
\noindent
Here, we rely on a formulation of~\cite{vub1} and \cite{vub2}, taking into account
recent measurements of the $B$ lifetime of $1.604 \pm 0.012\,\mbox{ps}\,$~\cite{pdg2004},
the $b$ quark mass in the kinetic mass scheme,  $m_b(1 \gev)=4.61 \pm 0.07 \gev$,
and other parameters of the heavy quark expansions~\cite{babarvcb}.
The first error represents the linear sum of the perturbative and non-perturbative
QCD corrections, the second error is due to the uncertainty in $m_b$.  An
overall  correction of 0.7\% is included to account for QED corrections.
The results for the four momentum intervals are presented in
Table~\ref{table:br2f}. They  are consistent with each other.
For the interval from 2.0 to 2.6~\gevc we quote as a preliminary result,
\begin{equation}
|V_{ub}|= 
(3.94 \pm 0.25_{exp} \pm 0.37_{f_u} \pm 0.19_{theory}) \times 10^{-3}.
\end{equation}
Here the first error represents the total experimental uncertainty, and the second refers to the
uncertainty on the determination of the fraction $f_u$ from the $B \ra X_s \gamma$
decays (taken from the CLEO analysis), and the third combines the stated
theoretical uncertainties
in the extraction of $|V_{ub}|$ from the branching ratio.
No additional uncertainty due to the theoretical assumption of quark-hadron 
duality has been assigned.

Recently new preliminary determination of the shape function parameters
based on analysis of $b\to s\gamma$ spectrum by the Belle collaboration 
became available \cite{belle} with the
central values of $a=2.27$ and $m_b^{SF}=4.62$\gevcc\ .
The Belle shape-function parameters differ from those of CLEO by more than one
standard deviation. The results for $B\to X_u e\nu$ partial and total branching ratios
based on this new set of shape-function parameters and their errors  
are presented in the Table~\ref{table:br2f_belle}.
The quoted errors on $\Delta {\cal B}$ are statistical and systematic, respectively.
The total error on $f_u$ includes the
statistical and systematic errors of the shape-function measurement and the
uncertainty of the theoretical calculation of $f_u$ from $b\to s\gamma$
shape function, estimated as in \cite{cleo2}. The errors on $\cal B$
and $|V_{ub}|$ are the total error from $\Delta {\cal B}$ measurement
and the error from $f_u$, respectively. The last error in  $|V_{ub}|$ is
the theoretical uncertainty of the translation from $\cal B$ to $|V_{ub}|$.

\begin{table}[h]
\caption{
Preliminary results on the partial ($\Delta {\cal B}$) and the total branching
fraction (${\cal B}$) for inclusive $B \to X_u e \nu$ decays for four momentum intervals. 
The recent Belle measurement of $b \to s \gamma$ spectrum was used  \cite{belle}. 
}
\label{table:br2f_belle}
{\small
\begin{center}
\begin{tabular}{ccccc} \hline\hline
$\Delta p $ (\gevc) & $\Delta {\cal B}\,(10^{-3})$ & $f_u(\Delta p)$     & ${\cal 
B}\,(10^{-3})$ & $|V_{ub}|\,(10^{-3})$ \\ \hline
$2.0\div 2.6$ & $0.531\pm 0.032\pm 0.049$ & $0.246\pm 0.031$ & $2.16\pm
0.24\pm 0.27$ & $4.40\pm 0.24\pm 0.28\pm 0.21$ \\
$2.1\div 2.6$ & $0.381\pm 0.020\pm 0.034$ & $0.174\pm 0.026$ & $2.19\pm
0.23\pm 0.33$ & $4.44\pm 0.23\pm 0.33\pm 0.21$ \\
$2.2\div 2.6$ & $0.245\pm 0.011\pm 0.020$ & $0.110\pm 0.022$ & $2.23\pm
0.21\pm 0.45$ & $4.47\pm 0.21\pm 0.45\pm 0.22$ \\
$2.3\div 2.6$ & $0.153\pm 0.006\pm 0.011$ & $0.058\pm 0.017$ & $2.64\pm
0.22\pm 0.77$ & $4.87\pm 0.20\pm 0.71\pm 0.23$ \\
\hline \hline
\end{tabular}
\end{center}
}
\end{table}

In conclusion, we have a preliminary measurement of the differential
lepton spectrum for charmless semileptonic $B$ decays above 2.0 \gevc,
and have extracted the CKM matrix element $|V_{ub}|$ with improved experimental
accuracy.  Further improvements to this measurement are expected from new
measurements of the inclusive photon spectrum from $B \ra X_s \gamma$ decays
and further advances in our theoretical understanding of the shape functions,
their relation to the parton-level lepton spectrum, and effects of the fragmentation
of the $s$ and $u$ quarks.

\section{ACKNOWLEDGMENTS}
\label{sec:acknowledgments}
We would like to thank the CLEO and Belle Collaborations for
providing detailed information on the extraction of 
the shape function parameters from the photon spectrum
in $b\to s \gamma$ transition.

We are grateful for the 
extraordinary contributions of our \pep2\ colleagues in
achieving the excellent luminosity and machine conditions
that have made this work possible.
The success of this project also relies critically on the 
expertise and dedication of the computing organizations that 
support \babar.
The collaborating institutions wish to thank 
SLAC for its support and the kind hospitality extended to them. 
This work is supported by the
US Department of Energy
and National Science Foundation, the
Natural Sciences and Engineering Research Council (Canada),
Institute of High Energy Physics (China), the
Commissariat \`a l'Energie Atomique and
Institut National de Physique Nucl\'eaire et de Physique des Particules
(France), the
Bundesministerium f\"ur Bildung und Forschung and
Deutsche Forschungsgemeinschaft
(Germany), the
Istituto Nazionale di Fisica Nucleare (Italy),
the Foundation for Fundamental Research on Matter (The Netherlands),
the Research Council of Norway, the
Ministry of Science and Technology of the Russian Federation, and the
Particle Physics and Astronomy Research Council (United Kingdom). 
Individuals have received support from 
CONACyT (Mexico),
the A. P. Sloan Foundation, 
the Research Corporation,
and the Alexander von Humboldt Foundation.

\end{document}